\documentclass[a4paper]{jpconf}
\usepackage{graphicx}
\usepackage[applemac]{inputenc}
\makeatletter
  \def\text#1{%
    \relax
    \ifmmode
      \mathchoice
	{\hbox{{\everymath{\displaystyle     }#1}}}%
	{\hbox{{\everymath{\textstyle        }#1}}}%
	{\hbox{{\everymath{\scriptstyle      }\let\f@size\sf@size\selectfont#1}}}%
	{\hbox{{\everymath{\scriptscriptstyle}\let\f@size\ssf@size\selectfont#1}}}%
      \glb@settings
    \else
      \mbox{#1}%
    \fi
  }
\makeatother
\begin{document}
\title{Two-dimensional frustrated spin systems in high magnetic
fields}

\author{B Schmidt$^{1}$, N Shannon$^{2}$ and P Thalmeier$^{1}$}

\address{$^{1}$ Max-Planck-Institut für Chemische Physik fester 
Stoffe, 01187 Dresden, Germany}

\address{$^{2}$ H H Wills Physics Laboratory, Tyndall Avenue, Bristol
BS8 1TL, United Kingdom}

\ead{bs@cpfs.mpg.de}

\begin{abstract}
    We discuss our numerical results on the properties of the $S=1/2$
    frustrated $J_{1}$-$J_{2}$ Heisenberg model on a square lattice as
    a function of temperature and frustration angle
    $\phi=\tan^{-1}(J_{2}/J_{1})$ in an applied magnetic field.  We
    cover the full phase diagram of the model in the range
    $-\pi\le\phi\le\pi$.  The discussion includes the parameter
    dependence of the saturation field itself, and addresses the
    instabilities associated with it.  We also discuss the
    magnetocaloric effect of the model and show how it can be used to
    uniquely determine the effective interaction constants of the
    compounds which were investigated experimentally.
\end{abstract}

\section{Introduction}

The ground-state and low-temperature properties of low-dimensional
frustrated spin systems are of great interest due to their unusual and
sometimes unexpected temperature and magnetic-field dependence.  The
origin for this lies in the enhanced quantum fluctuations which
determine their behaviour.  A prime example for such a system is the
spin-$1/2$ frustrated $J_{1}$-$J_{2}$ Heisenberg model on a square
lattice, being one of the most intensively studied models for spin
systems.  It has a Hamiltonian of the form
\begin{equation}
    H = J_{1}\sum_{\left\langle ij\right\rangle_{1}}{\bf S}_{i}{\bf 
    S}_{j} + J_{2}\sum_{\left\langle ij\right\rangle_{2}}{\bf S}_{i}{\bf 
    S}_{j},
    \label{eqn:hamiltonian}
\end{equation}
where $J_{1}$ denotes the exchange interaction between neighbouring
spins, and $J_{2}$ labels the exchange interaction along the diagonals
between next nearest neighbours.

In order to facilitate the description of the model and its 
thermodynamics, it is convenient to introduce two parameters to 
characterise it, namely an overall energy scale $J_{\text c}$ and a 
frustration angle $\phi$ defined by
\begin{equation}
    J_{\text c}=\sqrt{J_{1}^{2}+J_{2}^{2}}\quad\mbox{and}\quad
    \phi=\tan^{-1}\left(J_{2}/J_{1}\right),
    \label{eqn:phi}
\end{equation}
such that $J_{1}=J_{\text c}\cos\phi$, and $J_{2}=J_{\text
c}\sin\phi$.  Except where stated otherwise, the numerical results
discussed in the following are all obtained using the
finite-temperature Lanczos method~\cite{shannon:04} on a 20-site
square with periodic boundary conditions.

\section{Phase diagram and experiments}
\begin{figure}
    \centering
    \includegraphics[height=.4\columnwidth]{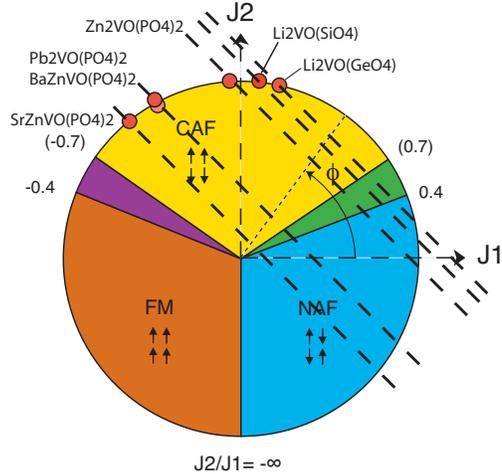}
    \caption{Schematic phase diagram of the $J_{1}$-$J_{2}$ model.
    The boundary between FM and NAF phase is the line $J_{1}=0$,
    $J_{2}<0$.  Numbers are ratios of exchange couplings $J_{2}/J_{1}$
    where zero point fluctuations destroy the relevant order
    parameter~\protect\cite{shannon:04,shannon:06}.  The dashed lines
    labeled by the chemical formulae for the experimentally known
    ``$J_{1}$-$J_{2}$ compounds'' are lines of constant
    $\Theta_{\text{CW}}=(J_{1}+J_{2})/k_{\text B}$ as determined from
    $H=0$ heat capacity and magnetic susceptibility
    measurements~\protect\cite{kaul:05,kini:06}.}
    \label{fig:phasediagram}
\end{figure}
Figure~\ref{fig:phasediagram} shows the phase diagram of the model,
together with some extra lines referring to the compounds which were
investigated experimentally.  At a classical level, the phase diagram
has three main phases: For $-\pi/2<\phi<\tan^{-1}(1/2)$, the N\'{e}el
antiferromagnet (NAF) characterised by an ordering vector ${\bf
Q}=(\pi,\pi)$, a ferromagnetic phase (FM) for
$\pi-\tan^{-1}(1/2)<\phi<-\pi/2$, and a collinear phase (CAF) with
${\bf Q}=(\pi,0)$ or $(0,\pi)$ for
$\tan^{-1}(1/2)<\phi<\pi-\tan^{-1}(1/2)$.  At the boundary between CAF
and NAF, the Ising domain wall energy between the states vanishes, and
arbitrary stripes of the two phases can be formed.  At the CAF/FM
boundary, coplanar spiral states with ${\bf q} = (2\pi n/m, 0)$, where
$\{n,m\}$ are integers, all become degenerate.
Quantum fluctuations lift these degeneracies and lead to disordered
regions of finite width at these two boundaries having continuous
classical degeneracies.  A more detailed discussion of the properties
of the model at zero magnetic field is given in
References~\cite{shannon:04,shannon:06} and the references cited
therein.

The $J_{1}$-$J_{2}$ model is appropriate for the description of a
number of quasi two-dimensional vanadium compounds.  They can be
grouped in two families of type Li$_{2}$VO$X$O$_{4}$ ($X = \text{Si,
Ge}$), and $AA'$VO(PO$_{4}$)$_{2}$ ($A,A'=\text{Pb, Zn, Sr, Ba}$),
respectively~\cite{kaul:05,kini:06}.  All compounds feature V$^{4+}$
ions with $S=1/2$ surrounded by oxygen polyhedra, forming layers of
$J_{1}$--$J_{2}$ square lattices which are weakly coupled in the third
spatial dimension.


Thermodynamic coefficients have been measured to determine the
effective interaction constants~\cite{kaul:05,kini:06}.  For zero
magnetic field, this is not completely possible.  An ambiguity remains
when fitting the experimental $\chi(T)$ and $C_{V}(T)$ using a
high-temperature series expansion: We always have {\em two\/} possible
points $\phi=\phi_{\pm}$ in the phase diagram symmetric around
$\phi=\pi/4$.  This situation does {\em not\/} change qualitatively
when including higher-order terms in the expansion used for the
fits~\cite{misguich:03}.

\section{Finite fields}

Fortunately, all $J_{1}$-$J_{2}$ compounds known so far have values
$J_{\text c}={\cal O}(10 \text K)$, which corresponds to a
magnetic-field range accessible in pulsed high-magnetic field
experiments.

\begin{figure}
    \centering
    \hfill
    \includegraphics[height=0.32\columnwidth]{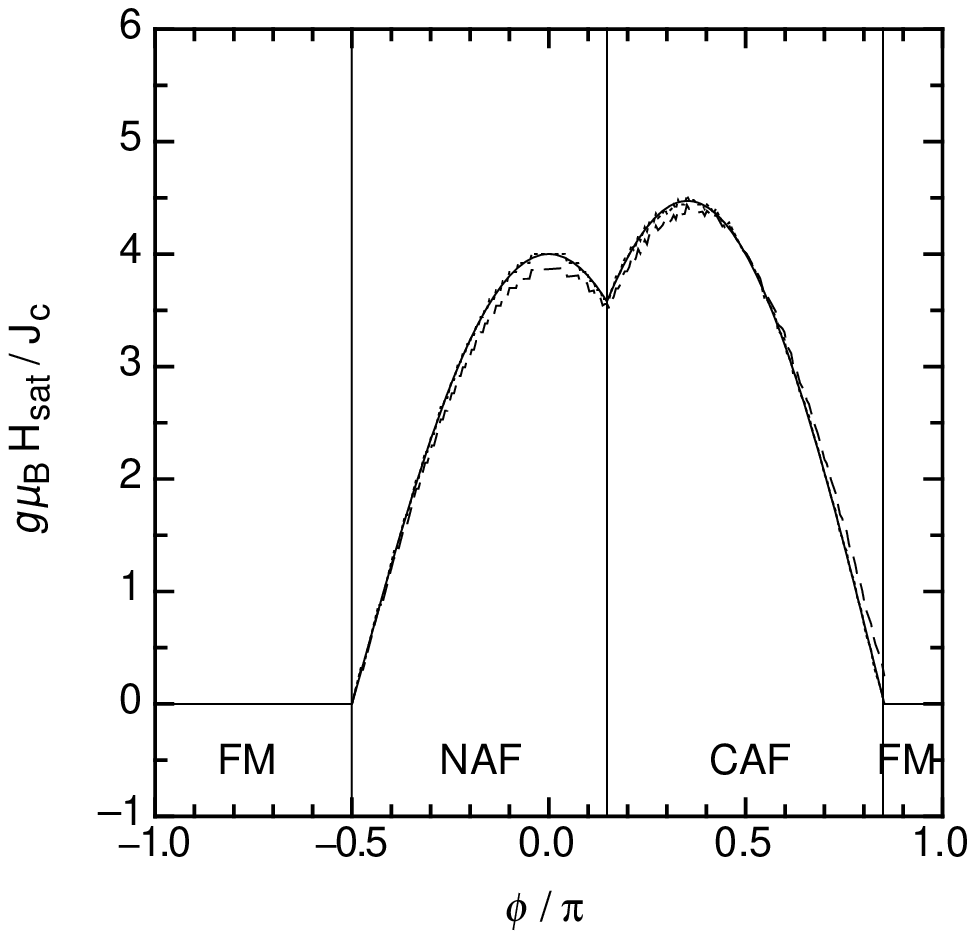}
    \hfill
    \includegraphics[height=0.32\columnwidth]{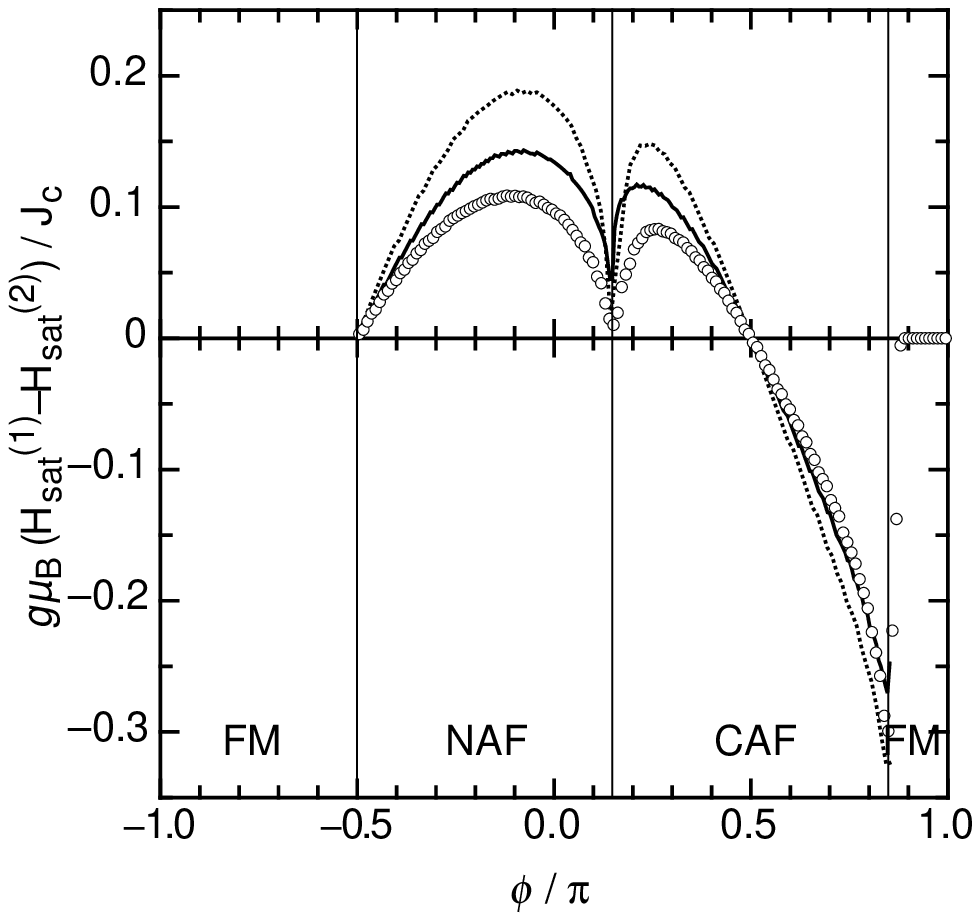}
    \hfill \null \caption{Left: One-magnon instability (solid line)
    and two-magnon instability (dashed line) field $H$ of the fully
    polarized state of the $J_{1}$-$J_{2}$ model on a 20-site square
    as a function of the frustration angle
    $\phi=\tan^{-1}(J_{2}/J_{1})$.  The one-magnon instability field
    corresponds to the saturation field of the magnetisation obtained
    by linear spinwave theory.  The energy unit is $J_{\rm
    c}=\sqrt{J_{1}^{2}+J_{2}^{2}}$.  Right: Difference between the
    $\Delta S=1$ and $\Delta S=2$ instabilities as a function of the
    frustration angle for a cluster with 16 sites (dotted line), 20
    sites (solid line), and 24 sites (open circles).}
    \label{fig:saturation}
\end{figure}
We have calculated the saturation field of the magnetisation as a
function of the frustration angle $\phi$ and the magnetic field $H$.
The left-hand side of Figure~\ref{fig:saturation} shows the result for
a 20-site cluster with periodic boundary conditions as well as the
saturation field obtained from linear spinwave theory.  We define the
$k^{\text{th}}$ saturation field as the instability of the fully
polarised (saturated) state against an excitation with $k$ magnons,
\begin{equation}
    g\mu_{\rm B}H_{\rm sat}^{(k)}=
    \frac{1}{k}\left[E\left(S_{\rm 
    tot}^{z}=\frac{N}{2}\right)-E\left(S_{\rm 
    tot}^{z}=\frac{N}{2}-k\right)\right],
    \label{eqn:hsatk}
\end{equation}
where $E$ denotes the ground-state energy per site of the cluster of
size $N$, and $S_{\rm tot}^{z}$ is the component of the total spin of
the system in the direction of the magnetic field $H$.  In this
notation, the thermodynamic saturation field of the magnetisation is
given by $H_{\text{sat}}=\max_{k}\left(H_{\rm
sat}^{(k)}\right)$.  The result from linear spinwave theory for the
infinitely large system reads
\begin{equation}
    g\mu_{\rm B}H_{\rm sat}^{({\rm cl})}=
    zSJ_{\rm c}\left[\cos\phi\left(1-\frac{1}{2}\left(\cos Q_{x}+\cos 
    Q_{y}\right)\right)+
    \sin\phi\left(\vphantom{\frac{1}{2}}1-\cos Q_{x}\cos Q_{y}\right)\right]
    \label{eqn:hsatcl}
\end{equation}
with $z=4$ and $S=1/2$, and is displayed as solid line in
Figure~\ref{fig:saturation} (left).  It is in perfect agreement with
the numerical result for $H_{\rm sat}^{(1)}$.

For values $-\pi/2<\phi<\pi/2$ ($J_{1}>0$), the saturation field is
determined by the $\Delta S=1$ instability.  Inside the collinear
phase, at the point $\phi=\pi/2$, $J_{1}$ vanishes, and the lattice
decouples into two sublattices.  Therefore, the $\Delta S=1$ and
$\Delta S=2$ spin flip energies must have the same value.  For
ferromagnetic $J_{1}$, for the finite-size clusters considered, the
$\Delta S=2$ instability determines the saturation field until the
point $\phi\approx0.9\pi$ where the ferromagnetic phase is reached.
This is seen from Figure~\ref{fig:saturation} (right), where we have
plotted the difference $H_{\text{sat}}^{(1)}-H_{\text{sat}}^{(2)}$ for
three different cluster sizes.  While the overall size of
$H_{\text{sat}}^{(1)}-H_{\text{sat}}^{(2)}$ decreases monotonically
with the size of the cluster, at least in the NAF and CAF phases, the
crossover described above happens always exactly at $J_{1}=0$
independent of the cluster size.

\section{Magnetocaloric effect}

\begin{figure}
    \centering
    \hfill
    \includegraphics[height=0.34\columnwidth]{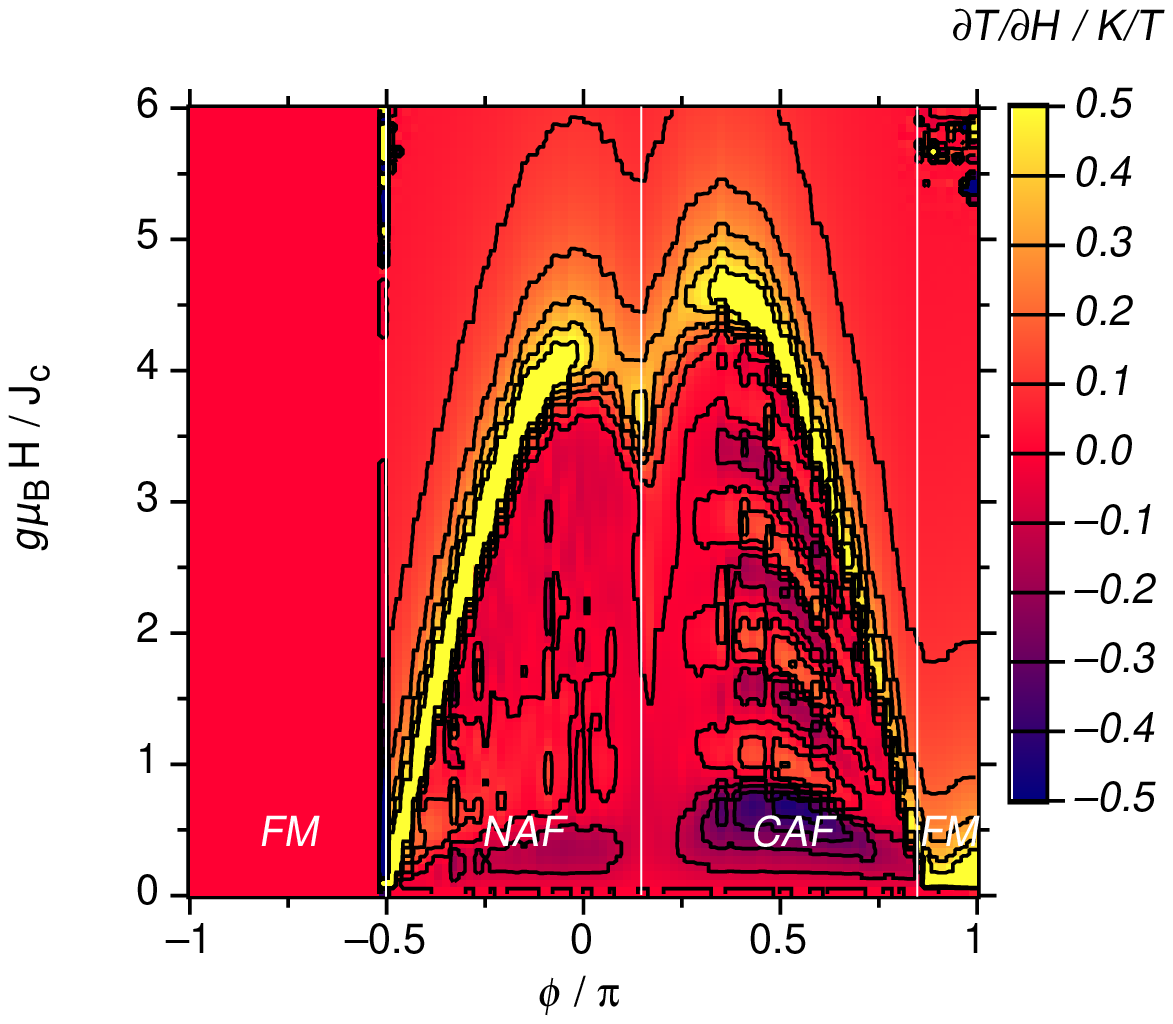}
    \hfill
    \includegraphics[height=0.32\columnwidth]{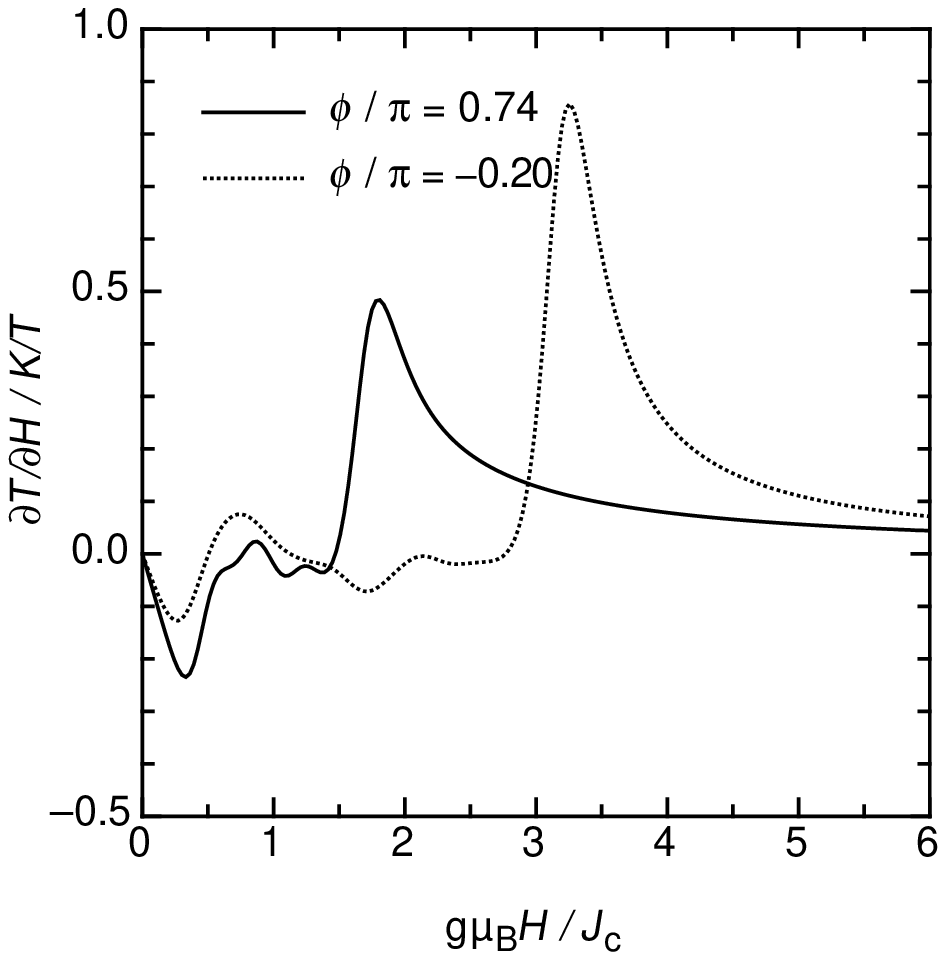}
    \hfill
    \null
    \caption{Left: $H$-$\phi$ contour plot of the MCE of the
    $J_{1}$-$J_{2}$ model at $T=0.2J_{\text c}/k_{\text B}$ on a
    20-site square with periodic boundary conditions.  The calculation
    was done on a grid of $330\times200$ points.  Right: Field
    dependence of the MCE for those two particular values of
    $\phi=\phi_{\pm}$ determined for the compound
    SrZnVO(PO$_4$)$_2$~\protect\cite{kaul:05}, again at a temperature
    $T=0.2J_{\text c}/k_{\text B}$. Note the large difference of the 
    maximum positions.}
    \label{fig:mce}
\end{figure}
The magnetocaloric effect (MCE) is the adiabatic temperature change of
a sample by applying a magnetic field, given by

\begin{equation}
    \left(\frac{\partial T}{\partial H}\right)_{S}=
    -\left(\frac{\partial S}{\partial H}\right)_{T}\left/
    \left(\frac{\partial S}{\partial T}\right)_{H}\right.=
    -\frac{T(H)}{C_{V}(H,T)}\left(\frac{\partial M(H,T)}{\partial 
    T}\right)_{H},
    \label{eqn:mce}
\end{equation}
where $S$ is the entropy, and $M$ the magnetisation.  We have
calculated the MCE on a 20-site square as a function of $T$, $H$, and
$\phi$.  Figure~\ref{fig:mce} (left) shows the MCE for fixed
temperature $T=0.2J_{\text c}/k_{\text B}$.  Its main feature is the
pronounced anomaly near the saturation field where the fully polarised
state is reached, corresponding to the sudden entropy change taking
place upon reaching saturation.  The saturation magnetisation is
reached slightly above the maximum position of the MCE. The
oscillating structure of the MCE below $H_{\text{sat}}$ in the CAF
phase is a finite-size effect which we expect to vanish in the
thermodynamic limit.  Note that in the Néel as well as in the
collinear phase, the MCE is for some range of fields {\em negative\/},
indicating that the sample will {\em cool down\/} with increasing
field, opposite to what is expected for simple paramagnets.

The enhanced quantum fluctuations due to the frustration are evidenced
by an enhanced magnetocaloric effect in the disordered region between
the CAF and the FM phases, and close to the region between the CAF and
the NAF phases.

The right-hand side of Figure~\ref{fig:mce} shows two vertical cuts
through the left-hand side plot for those values $\phi=\phi_{\pm}$
which have been determined experimentally from zero-field heat
capacity and susceptibility measurements of the compound
SrZnVO(PO$_4$)$_2$~\protect\cite{kaul:05}.  In contrast to the
zero-field data, the two maximum positions of the MCE anomalies differ
by roughly a factor two.

The maximum saturation field of the $J_{1}$-$J_{2}$ model is reached
at $\phi_{\text{max}}/\pi=\tan^{-1}(2)/\pi\approx0.35$ with
$g\mu_{\text B}H_{\text{sat}}^{\text{max}}=4.5J_{\text c}$.
Since the available compounds all have values $J_{\text
c}={\cal O}(10\,{\text K})$, their saturation fields have values
$H_{\text{sat}}<40\,\text T$, assuming $g=2$.  These fields are
accessible; therefore, a measurement of the field dependence of the
magnetocaloric effect can help to clarify the nature of the ground
state of the $J_{1}$-$J_{2}$ compounds discussed here.

\appendix

\section*{References}

\bibliographystyle{unsrt}
\bibliography{thp68}

\end{document}